\documentclass[page-classic]{epl2} 

\begin{document}

\title{     Orbital correlations in monolayer manganites --- \\
            From spin $t$--$J$ model to orbital $t$--$J$ model
       \thanks{This work is dedicated to Professor J\'ozef Spa\l{}ek 
               on the occasion of his $60^{th}$ birthday.}}

\shorttitle{ Orbital correlations in monolayer manganites } 

\author{     Maria Daghofer\inst{1} \and Andrzej M. Ole\'s \inst{1,2}}
\shortauthor{M. Daghofer and A. M. Ole\'s}

\institute{                    
  \inst{1} Max-Planck-Institut f\"{u}r Festk\"{o}rperforschung,\\
           Heisenbergstrasse 1, D-70569 Stuttgart, Germany\\
  \inst{2} Marian Smoluchowski Institute of Physics, Jagellonian 
           University,\\ Reymonta 4, PL-30059 Krak\'ow, Poland
}

\abstract{
On the example of monolayer manganites we show that the theoretical 
ideas developed long ago along the derivation of the spin $t$--$J$ 
model from the Hubbard model are nowadays very helpful in strongly 
correlated oxides with partly filled degenerate orbitals. We analyze 
a realistic orbital $t$--$J$ model for $e_g$ electrons in 
La$_{1-x}$Sr$_{1+x}$MnO$_4$ monolayer manganites, and discuss the 
evolution of spin and orbital correlations under increasing doping 
by performing exact diagonalization of finite clusters, with 
electronic kinetic energy determined self-consistently at finite 
temperature by classical Monte Carlo for $t_{2g}$ spins. Several 
experimental results are reproduced.
}

\maketitle

\section{Spin $t$--$J$ model}

A new perspective to treat strongly correlated electrons was opened by 
the derivation of the spin $t$--$J$ model three decades ago 
\cite{Cha77}. We argue that this derivation opened new routes to 
understand the complex magnetic and superconducting instabilities in 
the Hubbard model, and proved to be a very useful paradigm in strongly 
correlated electron systems. 
At that time it was already realized that the Hubbard model,
\begin{equation}
\label{Hub}
{\cal H}_0=-t\sum_{ij\sigma}a_{i\sigma}^{\dagger}a_{j\sigma}^{}
+U\sum_i n_{i\uparrow}n_{i\downarrow}\;,
\end{equation}
where $a_{i\sigma}^{\dagger}$ is an electron creation operator in spin 
state $\sigma=\uparrow,\downarrow$ at site $i$ and $U$ is local Coulomb 
interaction, is perfectly designed to describe electron localization in 
a Mott insulator, when the average filling is $n=1$ electron per site, 
and the Coulomb repulsion is large, i.e., $(t/U)\ll 1$. 
In this case the low-energy physics is determined by the effective 
superexchange interaction between localized electron states, which is 
antiferromagnetic (AF) due to the Pauli principle. In this limit the 
energy is determined by the nearest neighbor spin correlations and by 
higher order correction terms in powers of $(t/U)$ \cite{Cha78}. 

However, when a Mott insulator is doped by holes and electron density 
is reduced to $n=1-x$, the kinetic energy competes with the AF 
superexchange and may lead to novel physical phenomena \cite{Spa07}. 
Usually one considers then the effective {\it spin\/} $t$--$J$ 
{\it model\/}. 
\begin{equation}
\label{tJ}
{\cal H}_{tJ}=
-t\sum_{ij\sigma}{\tilde a}_{i\sigma}^{\dagger}{\tilde a}_{j\sigma}^{}
+J\sum_{\langle ij\rangle}{\vec S}_i\cdot{\vec S}_j\;,
\end{equation}
derived first in Ref. \cite{Cha77},
where ${\vec S}_i$ are spin operators at each site occupied by a single
electron, and $J=4t^2/U$ is the superexchange constant. In contrast to 
Eq. (\ref{Hub}), the kinetic energy term $\propto t$ contains 
'constrained' fermion creation operators, 
${\tilde a}_{i\sigma}^{\dagger}=a_{i\sigma}^{\dagger}(1-n_{i\bar{\sigma}})$
with $\bar{\sigma}=-\sigma$, 
and describes processes restricted to the subspace without doubly 
occupied sites. The lowest kinetic energy is obtained when spins are 
aligned in a ferromagnetic (FM) configuration. This competition between 
the kinetic and magnetic energy is the driving force responsible for 
the evolution of the magnetic phase diagram under increasing hole 
doping \cite{Spa81}. 

It was also recognized already in 1977 \cite{Cha77} that the faithful 
low-energy representation of the Hubbard model is not the $t$--$J$ model
itself, but rather the $t$--$J$ model extended by the three-site term
which contribute to the hole dynamics away from half filling:
\begin{equation}
\label{H3-site}
{\cal H}_{\rm eff}=
-t\sum_{ij\sigma}{\tilde a}_{i\sigma}^{\dagger}{\tilde a}_{j\sigma}^{}
+J\sum_{\langle ij\rangle}{\vec S}_i\cdot{\vec S}_j      
-\frac{1}{4}J\sum_{ikj,\sigma\sigma'}
{\tilde a}_{i\sigma}^{\dagger}a_{k\sigma}^{}n_{k\bar{\sigma}}^{}
n_{k\bar{\sigma'}}^{}a_{k\sigma'}^{\dagger}{\tilde a}_{j\sigma'}^{}\;.
\end{equation}
In fact, there is no reason to neglect this last term in Eq. 
(\ref{H3-site}) as it is of the same order as the effective exchange 
interaction $J$ \cite{Cha77}, and plays an important role except when 
doping $x\to 0$. Indeed, the three-site term captures an important part 
of the dynamics near a Mott insulating state and is essential to give 
the correct increase of the Drude weight with doping $x$ \cite{Ste92}, 
while a constant Drude weight follows instead from the $t$--$J$ model. 
The three-site term is also necessary to reproduce the doping 
dependence of the spectral weight for the lower and upper Hubbard bands 
in the optical spectra, while the $t$--$J$ model misses most of the 
spectral weight transfer between the Hubbard subbands which occurs due 
to kinetic processes \cite{Esk94}. For the same reason, only the 
complete form of the effective Hamiltonian Eq. (\ref{H3-site}) is 
appropriate to analyze yet another type of order in the regime of 
large $U$ --- the superconducting phase. When the local pair operators, 
$b_{ij}=
({\tilde a}_{i\uparrow}^{\dagger}{\tilde a}_{i\downarrow}^{\dagger}
-{\tilde a}_{i\downarrow}^{\dagger}{\tilde a}_{i\uparrow}^{\dagger})
/\sqrt{2}$,
are introduced \cite{Spa88} for bond $\langle ij\rangle$, the effective 
Hamiltonian (\ref{H3-site}) describes correctly the dynamics of local 
pairs in a strongly correlated system. The three-site term gives an
important contribution which favors the superconducting states over 
the magnetic ones, and could explain the observed increase of the 
critical temperature in the high $T_c$ superconductors \cite{Spa88}.
In contrast, when a propagation of a single hole in an AF background is 
analyzed, the $t$--$J$ model (\ref{tJ}) is sufficient and gives new 
features in the electronic structure, such as quasiparticle band and 
incoherent part of the spectral density, which result from the coupling 
of a moving hole to spin excitations \cite{Mar91}, while the three-site 
term leads only to quantitative corrections \cite{vSz90} which become
important only at large $J\sim t$ \cite{Bal95}. Therefore, the $t$--$J$ 
model is also extensively used for interpreting the data of 
photoemission experiments for high $T_c$ superconductors at low hole 
doping \cite{Dam03}. In recent years the $t$--$J$ model continues to 
play an important role in the theoretical studies of high $T_c$ 
superconductors, and serves to identify novel complex types of order 
\cite{Rac06}.

{\it Inter alia\/}, the idea standing behind the $t$--$J$ model is 
extremely helpful to describe the doped correlated insulators with 
orbital degrees of freedom. One decade after the discovery of high $T_c$ 
superconductivity the interest in strongly correlated systems has to a 
large extent moved to the systems with partly filled degenerate 
orbitals, where electron localization and magnetic interactions are rich 
and lead to complex phase diagrams \cite{Spa98}. In the last 
decade of the last century it was also realized that carriers in doped 
Mott-Hubbard insulators bind to $d$--$d$ excitations which gives rise 
to rather general and complex models of $t$--$J$ variety \cite{Zaa93}.
Both the derivation and theoretical treatment of such modes are a
challenge in the modern theory of strongly correlated electrons. For 
instance, internal degrees of freedom of a moving carrier can change 
completely the behavior known from the spin $t$--$J$ model, where a 
singlet hole is responsible for coherent quasiparticles. In fact, a 
moving carrier with either orbital flavor or large spin (or both) leaves 
behind a path with its own history. Perhaps the simplest example of 
this behavior is provided by a triplet hole moving in an AF background 
\cite{Zaa92}, with its completely incoherent motion and suppressed 
tendency towards FM polarons known from the singlet case \cite{Spa81}.  

Yet, there are also cases where the quasiparticle states are obtained 
by a coupling of the moving singlet hole to orbital excitations. 
The {\it orbital\/} $t$--$J$ {\it model\/} \cite{vdB00} was introduced 
in a close analogy to Eq. (\ref{tJ}), and describes the motion of a hole 
in a FM plane of LaMnO$_3$. The orbital superexchange $J$ follows again 
from virtual charge excitations in the regime of $U\gg t$, which applies
to the realistic parameters of LaMnO$_3$ \cite{Fei99}. As in the spin
$t$--$J$ model \cite{Ste92}, the three-site term plays here an important 
role near the half filling (for weakly doped manganites), and is crucial
to reproduce the correct behavior of the Drude weight at increasing 
doping $x$ \cite{Hor99}. Nevertheless, we neglect the three-site terms 
as they play almost no role at high doping $x\simeq 0.5$ which is of 
interest in the present study.

For a FM plane it is allowed to decouple spin and orbital operators
following the mean field procedure, although in general this would lead 
to incorrect results. In low-spin states the spin-orbital entanglement 
\cite{Ole06} generates novel ground states which do not follow from the 
classical expectations, such as Goodenough-Kanamori rules \cite{Goode}. 
The orbital order (OO) in an undoped 
insulator is already a challenge as the orbitals are directional, the 
orbital (superexchange) interactions depend on the bond direction 
\cite{Zaa93}, and their symmetry is thus cubic, i.e., much lower than 
the SU(2) symmetry in the spin case. As a result, the coherent 
quasiparticle states for a single hole moving in a Mott insulator with 
alternating orbital (AO) order have somewhat lower weight than in spin 
case \cite{vdB00}, and depend also on the type of the occupied orbitals.

Only in the last decade it has been recognized that the Jahn-Teller (JT) 
interactions and the superexchange which occurs due to $e_g$ electron 
excitations in the regime of large on-site Coulomb interaction $U$, 
support each other \cite{Fei99} and both are necessary to explain the 
magnetic and optical properties of undoped LaMnO$_3$ \cite{Ole05}. 
In doped manganites, like in La$_{1-x}$Sr$_{x}$MnO$_3$, the average 
density of $e_g$ electrons $n=1-x$ is tuned by doping, and the paradigm 
of the orbital $t$--$J$ model may help to understand the actual orbital 
correlations in various doping regimes. 
In the present paper we follow this idea and analyze the orbital 
correlations in monolayer La$_{1-x}$Sr$_{1+x}$MnO$_4$ manganites using 
the orbital $t$--$J$ model. The realistic model is richer than that of 
Ref. \cite{vdB00}, and contains also the JT term \cite{Dag04}, as well
as the crystal field term which follows from the two-dimensional (2D) 
geometry and lifts the degeneracy of $e_g$ orbitals. The orbital 
$t$--$J$ model constructed in this way includes all the terms which 
control the magnetic correlations in doped manganites \cite{Feh04}.

The monolayer manganites are the subject of intense recent experimental 
research. The undoped monolayer LaSrMnO$_4$ compound has the same 
magnetic structure as K$_2$NiF$_4$, with a 2D $G$-type AF ($G$-AF) order 
\cite{Sen05}. At $x=0.5$ the AF order is CE-type, as deduced from the 
x-ray and neutron scattering data for the structural and magnetic 
correlations \cite{Lar05}, as well as from the neutron measurements of 
magnetic excitations \cite{Bra06}. One expects that the orbital 
correlations which follow from strong correlations and the JT 
interactions play an important role in stabilizing this phase \cite{Lee06}. 
Furthermore, unlike in La$_{1-x}$Sr$_{x}$MnO$_3$, no FM metallic phase 
was found in doped monolayer La$_{1-x}$Sr$_{1+x}$MnO$_4$ compounds, but 
instead short-range magnetic correlations \cite{Sen05,Lar05} indicate 
that magnetic interactions are frustrated. This behavior is puzzling 
and will be addressed in the present study.

\section{Orbital $t$--$J$ model}

We consider the effective {\it orbital\/} $t$--$J$ {\it model\/},
\begin{equation}
\label{Horb}
{\cal H}=H_t+H_J+H_z+H_{\rm JT}+H_{V}+H_{J'}\;,
\end{equation}
introduced here by generalizing the one-dimensional (1D) model of Ref. 
\cite{Dag04} to doped monolayer manganites. The first three terms 
are the same as the ones in the orbital $t$--$J$ model \cite{vdB00}, 
while the other three serve to describe the realistic situation of 
$e_g$ electrons in La$_{1-x}$Sr$_{1+x}$MnO$_4$ manganites. The hopping 
term $H_t$ for monolayer manganites,
\begin{eqnarray}
H_t = -\frac{1}{4}t
\sum_{\langle ij\rangle\parallel ab} u_{ij}
  \Big\{3{\tilde c}_{ix}^{\dagger}{\tilde c}_{jx}^{}
        +{\tilde c}_{iz}^{\dagger}{\tilde c}_{jz}^{}     
\mp\sqrt{3}
 ({\tilde c}_{ix}^{\dagger}{\tilde c}_{jz}^{}
 +{\tilde c}_{iz}^{\dagger}{\tilde c}_{jx}^{}) + \mathrm{h.c.}\Big\}\;.
\label{Ht}
\end{eqnarray}
acts in the restricted space without double occupancies, so 
the operator ${\tilde c}_{i\alpha}^{\dagger}$ creates an electron in 
$|\alpha\rangle=|x\rangle,|z\rangle$ state at site $i$ only when this 
site is initially empty, i.e., ${\tilde c}_{i\alpha}^{\dagger}=
c_{i\alpha}^{\dagger}(1-n_{i\bar{\alpha}})$ with $\bar{\alpha}$ 
standing for the orbital favor opposite to $\alpha$, similar to spin 
in Eq. (\ref{tJ}). We use here the conventional $e_g$ orbital basis:
\begin{equation}
\label{realorbs}
\textstyle{
|x\rangle\equiv \frac{1}{\sqrt{2}} (x^2-y^2)\;,
\hspace{1.2cm}
|z\rangle\equiv \frac{1}{\sqrt{6}}(3z^2-r^2)\;.}
\end{equation}
Here $t$ stands for an effective $(dd\sigma)$ element that originates
from two $d$--$p$ transitions involving the intermediate oxygen ion. 
Following the double exchange mechanism \cite{Dag04}, the hopping in 
Eq. (\ref{Ht}) is modulated by the quantity $u_{ij}$ which depends on 
the spin state. Here the core spins ${\vec S}_i$ are treated as 
classical vectors of length $S$, so their state is represented by two 
polar angles $\{\vartheta_i,\phi_i\}$ per site, and the scalar product 
on the bond $\langle ij\rangle$ is given by
\begin{equation}
\label{sij}
\langle{\vec S}_i\cdot{\vec S}_{j}\rangle=S^2\big(2|u_{ij}|^2-1\big)\;,
\end{equation}
where the actual dependence on the spins configuration gives
\begin{equation}
\label{uij}
\textstyle{
u_{ij}=
\cos\big(\frac{1}{2}\vartheta_i\big)\cos\big(\frac{1}{2}\vartheta_j\big)
+e^{i(\phi_j-\phi_i)}\,
\sin\big(\frac{1}{2}\vartheta_i\big)\sin\big(\frac{1}{2}\vartheta_j\big)
=e^{i\chi_{ij}}\,\cos\big(\frac{1}{2}\theta_{ij}\big)\;.}
\end{equation}
Therefore, in the classical approximation the hopping (\ref{Ht}) is 
parameterized by the angle $\theta_{ij}$ between the two involved spins 
at sites $i$ and $j$, and by the complex phase $\chi_{ij}$. 

The superexchange in La$_{1-x}$Sr$_{1+x}$MnO$_4$ is given by a 
superposition of several terms which originate from charge excitations 
by either $e_g$ or $t_{2g}$ electrons. The superexchange terms which 
follow from charge excitations by $e_g$ electrons are included in $H_J$. 
These terms favor either FM or AF spin order on a considered bond 
$\langle ij\rangle$, depending on the pair of occupied $e_g$ orbitals at 
both sites. The form of $H_J$ depends on the actual used orbital basis,
and it is more convenient to use the directional orbitals and the 
corresponding pseudospin operators,
\begin{equation}
\label{Tzeta}
T_{i}^{\zeta}=-\textstyle{\frac{1}{2}}\big(T_{i}^z
                                \mp\sqrt{3}T_{i}^x\big)\;,
\end{equation}
depend on the bond direction, with the sign $-$ ($+$) in Eq.
(\ref{Tzeta}) corresponding to $a$ ($b$) axis. The operators are
defined by orbital $T=1/2$ pseudospin operators,
\begin{equation}
\label{Tz}
T_{i}^{z}=\textstyle{\frac{1}{2}}\sigma^z_i
=\textstyle{\frac{1}{2}}\big({\tilde n}_{ix}-{\tilde n}_{iz}\big)\;, 
\hskip 1cm  
T_{i}^{x}=\textstyle{\frac{1}{2}}\sigma^x_i
=\textstyle{\frac{1}{2}}
\big({\tilde c}^\dagger_{ix}{\tilde c}^{\phantom{\dagger}}_{iz}
    +{\tilde c}^\dagger_{iz}{\tilde c}^{\phantom{\dagger}}_{ix}\big)\;,
\end{equation}
with two eigenstates of $T_{i}^{z}$, see Eq. (\ref{realorbs}). 

The form of $H_J$ used below was derived in Ref. \cite{Dag04} from the 
complete spin-orbital model \cite{Fei99}. It focuses on the orbital 
dynamics in the presence of spin fluctuations which tune orbital 
superexchange interactions. The spin operators on the bonds 
(Mn$^{3+}$--Mn$^{3+}$ or Mn$^{3+}$--Mn$^{4+}$) 
are replaced by their expectation values which involve classical 
$u_{ij}$ parameters given in Eq. (\ref{uij}). In the present case of 
a monolayer one finds the $e_g$ superexchange term \cite{Dag04},
\begin{eqnarray}
\label{HJ}
H_J&=& J\sum_{\langle ij \rangle\parallel ab}\Big\{
 \frac{1}{5} \Big(2|u_{ij}|^2+3\Big)\Big( 2T_{i}^{\zeta}T_{j}^{\zeta}
-\frac{1}{2} {\tilde n}_i{\tilde n}_{j} \Big) 
-\frac{9}{10}\big(1-|u_{ij}|^2\big)
            {\tilde n}_{i\zeta}{\tilde n}_{j\zeta}  \nonumber \\
&-&\big(1-|u_{ij}|^2\big)
\big[{\tilde n}_{i\zeta}(1-{\tilde n}_{j})
+(1-{\tilde n}_i){\tilde n}_{j\zeta}\big]\Big\}\;,
\end{eqnarray}
where electron number operator ${\tilde n}_{i\zeta}$ refers in each case 
to the directional $(3z^2-r^2)$-like orbital $|\zeta\rangle$ along a 
given bond $\langle ij \rangle$, e.g. $3x^2-r^2$ orbital along $a$ axis. 
The last term in Eq. (\ref{HJ}) stands for the Mn$^{3+}$--Mn$^{4+}$ 
superexchange and favors configurations of an occupied directional 
orbital next to a hole on the bond, where ${\tilde n}_i$ is the electron 
density operator in the restricted space. The superexchange constant 
$J=t^2/\bar{U}$ in Eq. (\ref{HJ}) is determined by the lowest excitation 
energy $\bar{U}$ to the high-spin $S=5/2$ state of $d^5$ configuration 
at the involved Mn$^{2+}$ ion \cite{Fei99}.

The orbital $t$--$J$ model \cite{vdB00} contains also a uniform crystal 
field splitting $E_z>0$ of $e_g$ orbitals which removes the orbital 
degeneracy in monolayer manganites, and is expected from the structural 
data \cite{Sen05}. Therefore, we use 
\begin{equation}
H_z=\frac{1}{2}E_z\sum_i ({\tilde n}_{ix}-{\tilde n}_{iz})\;,
\label{Hz}
\end{equation}
where ${\tilde n}_{ix}$ and ${\tilde n}_{ix}$ stand for the electron
density operators in the two $e_g$ orbitals. Again, as in the spin 
$t$--$J$ model, these density operators act in the restricted Hilbert 
space, with double occupancies projected out. If $E_z>0$, the 
$|z\rangle$ orbitals are favored as expected for the undoped 
LaSrMnO$_4$ \cite{Sen05}, and at $x=0.5$ doping in 
La$_{0.5}$Sr$_{1.5}$MnO$_4$ \cite{Wil05}.

The JT term $H_{\rm JT}$ includes two contributions: 
\begin{equation}
\label{HJT}
H_{\rm JT}=
 \kappa \sum_{\langle ij \rangle\parallel ab} T_i^{\zeta}T_j^{\zeta}
+\frac{1}{8}\kappa\sum_{\langle\langle ij\rangle\rangle\parallel ab}
 T_i^{\zeta}(1-{\tilde n}_{k})T_j^{\zeta}\;.
\end{equation}
First term stands for the orbital interaction between two Mn$^{3+}$ 
ions on neighboring sites --- it plays an important role in the undoped 
LaSrMnO$_4$ and at low hole doping. The second one was derived from 
oxygen distortions by Ba\l{}a, Horsch and Mack \cite{Bal04} and 
describes a second neighbor interaction along the bonds 
$\langle\langle ij\rangle\rangle$ parallel to either $a$ or $b$ axis; 
the coupling constant fixed here at $\frac{1}{8}\kappa$ was estimated 
from the distance dependence of the JT coupling. In addition, we also 
include the intersite Coulomb interaction between $e_g$ electrons, 
\begin{equation}
\label{HV}
H_{V}=V\sum_{\langle ij\rangle\parallel ab} {\tilde n}_i {\tilde n}_j
\end{equation}
-- it plays a role in the doped regime, separating $e_g$ electrons from
each other. 

Finally, 
the remaining part of the superexchange follows from charge excitations 
by $t_{2g}$ electrons between two Mn ions. Independently of their 
valence, the $t_{2g}$ orbitals are filled by three electrons which form
a core spin $S=3/2$ due to Hund's exchange $J_H$. Therefore, this part
of the superexchange is AF due to the Pauli principle, in analogy to  
the AF superexchange in the Hubbard model with nondegenerate orbitals.
One can verify by using the realistic parameters which describe the 
multiplet spectra of Mn ions that the different contributions to the
AF superexchange are of similar value \cite{Ole02}, so we describe the 
$t_{2g}$ superexchange by the Heisenberg Hamiltonian ($J'>0$),
\begin{equation}
\label{HJ'}
H_{J'}=J'\sum_{\langle ij\rangle}
\big({\vec S}_i\cdot{\vec S}_{j}-S^2\big)\;.
\end{equation}
For convenience, we consider here core spins ${\vec S}_i$ of unit 
length. Thereby the actual physical values of $S=3/2$ spins are 
compensated by a proper increase of $J'$. The classical state of core 
spins $\{{\vec S}_i\}$ is given by two polar angles 
$\{\vartheta_i,\phi_i\}$ per site, which determine the energy following 
Eq. (\ref{sij}), and tune the kinetic energy (\ref{Ht}) along each bond 
$\langle ij\rangle$.

The ground state of a monolayer obtained at a given doping $x$ is
characterized by electron density distribution which we quantify by 
average electron densities $\{n_x,n_z\}$ in orbitals $\alpha=x,z$.
The competing tendencies between FM and AF spin correlations at short
distances, coexisting with different types of OO, have been 
investigated by intersite spin and orbital correlations at distance 
${\vec r}$: 
\begin{equation}
\label{ss}
{\cal S}(\vec r)=\frac{1}{N}\sum_{i}
\big\langle {\vec S}_i\cdot{\vec S}_{i+\vec r}\big\rangle\;,\hskip .7cm
{\cal T}_{\theta}(\vec r)=\frac{1}{N}\sum_i
               \big\langle T_{i}(\theta)
                           T_{i+\vec r}(\theta)\big\rangle\;.
\end{equation}
The orbital operators are defined for a particular orbital basis 
at site $i$,
\begin{equation}
T_{i}(\theta)=T_i^z\,\cos\theta+T_i^x\,\sin\theta\;,
\label{oo}
\end{equation}
and the pseudospin operators $\{T_i^z,T_i^x\}$ are given
by Eqs.~(\ref{Tz}). For instance, $\theta=0$ corresponds
to $T_i^zT_{i+m}^z$, $\theta=\pi/2$ --- to $T_i^xT_{i+m}^x$, and
$\theta=2\pi/3$ --- to $T_i^{\zeta}T_{i+{\vec r}}^{\zeta}$, with
$\zeta$ standing for the directional orbital along $a$ axis.
The orbital correlations expected in undoped manganites are of AO
type on two sublattices, which suggests that the
orbital correlations ${\cal T}_{\theta}(\vec r)$ defined as in Eq.
(\ref{oo}) are predominantly negative for nearest neighbors. These
correlations were investigated along the (10) and (11) 
(and equivalent) directions in 2D clusters.

\section{Numerical results}

We employed two different numerical methods to investigate
finite clusters described by the orbital $t$--$J$ model (\ref{Horb}):
 ($i$) exact diagonalization at zero temperature ($T=0$) for fixed core
       spin configurations (FM, $G$-AF, $C$-AF, and CE phase), and
($ii$) Markov chain Monte Carlo (MC) at finite temperature $T>0$ for
       core spins \cite{Dag98}, combined with exact diagonalization for 
       the $e_g$ orbital problem.
The ground state at $T=0$ for representative electron fillings was
determined by solving the orbital $t$--$J$ model for several possible
types of spin order, using $\sqrt{8}\times\sqrt{8}$ and $4\times 4$ 
clusters with periodic boundary conditions. We solved the orbital 
problem obtained for the selected core spin configuration using the 
Lanczos algorithm. We then determined the global ground state by 
comparing the energies obtained for different magnetic phases.

At finite temperatures, we investigated the effective orbital $t$--$J$
model (\ref{Horb}) by making use of a combination of Markov chain MC 
algorithm for the core spins \cite{Dag98} with Lanczos diagonalization 
of the orbital problem. For each classical core spin configuration 
occurring in the MC runs, we defined the actual values of classical 
variables $\{u_{ij}\}$, and next solved the orbital model. In each case 
we obtained the free energy for that core spin configuration from the 
few lowest eigenstates, which was next used to decide acceptance in the 
MC runs. More details about the calculation method were given 
in Ref. \cite{Dag06}.

\begin{figure}[b!]
\includegraphics[width=6.5cm]{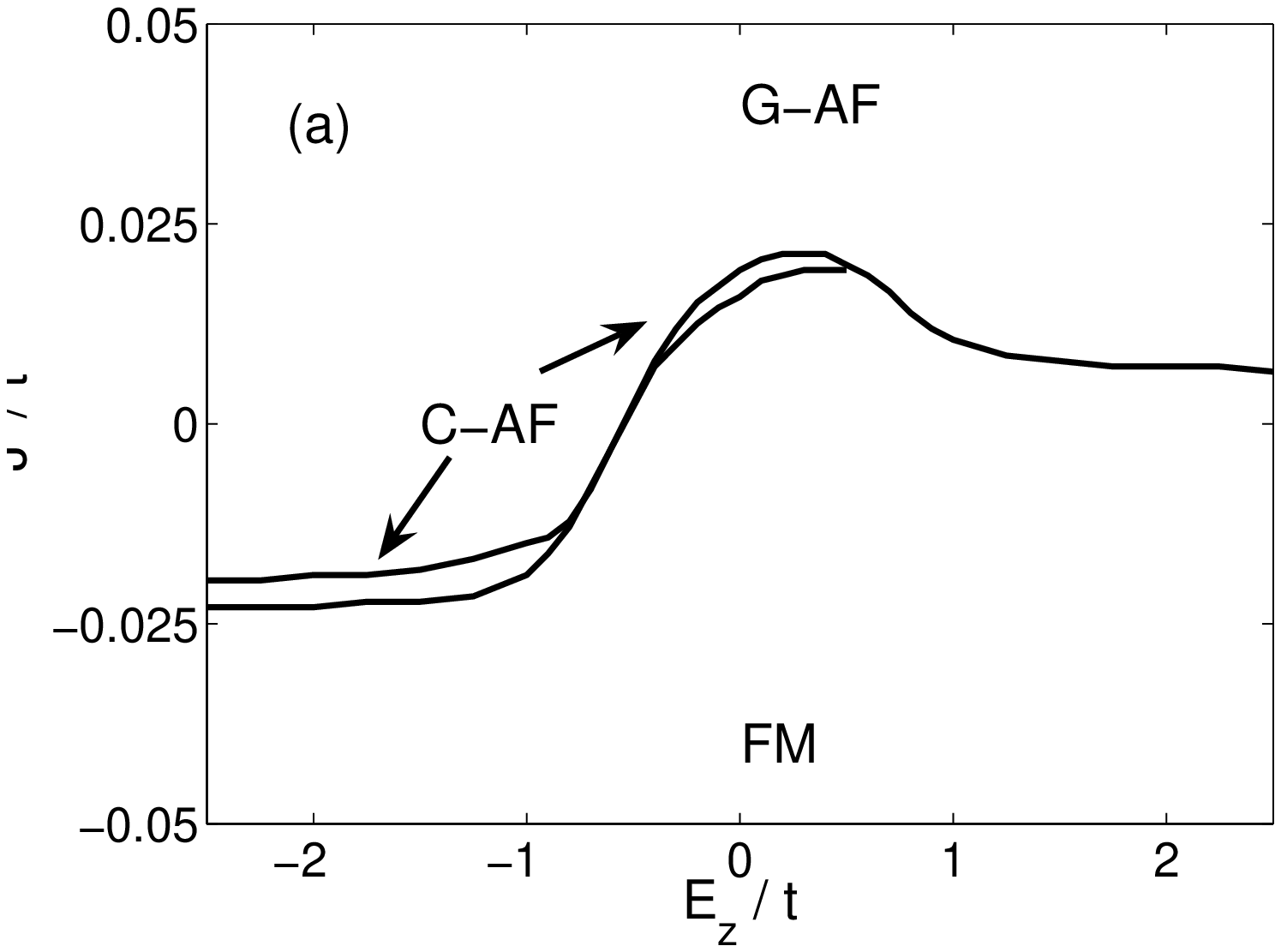}
\hskip .5cm
\includegraphics[width=6.5cm]{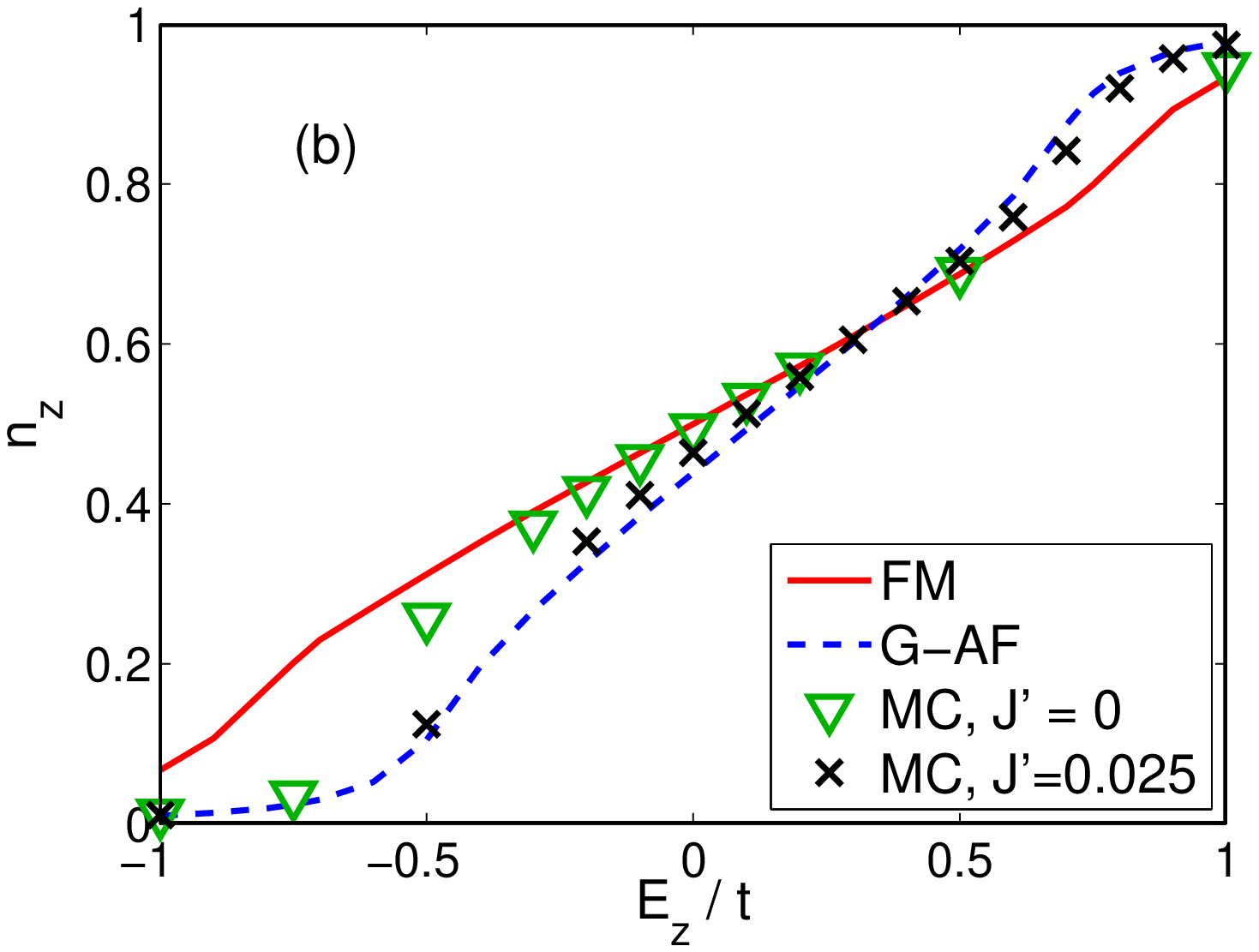}
\caption{(Color online)
Magnetic and orbital state for the undoped monolayer ($x=0$)
obtained with an $\sqrt{8}\times\sqrt{8}$ cluster:
(a) phase diagram with the stability regions of FM, $G$-AF, and $C$-AF 
    phases in the $(E_z,J')$ plane at $T=0$;
(b) electron density $n_z$ in out-of-plane $|z\rangle$ orbitals for 
    increasing value of the crystal field splitting $E_z$, as obtained 
    for the FM and AF ground states (solid and dashed line), as well as 
    the MC data for $\beta t=100$ and two values of $J'$. 
Parameters: $J=0.125t$, $\kappa=0.2t$.
}
\label{fig:x=0}
\end{figure}

We begin with the phase diagram obtained for the undoped LaSrMnO$_4$.
In absence of holes the hopping in Eq. (\ref{Ht}) is blocked, and 
the Coulomb interaction $V$ plays no role as it simply adds a constant 
energy for each bond. Therefore, the ground state for finite $J$ is 
determined by only three parameters: $E_z/J$, $\kappa/t$, and $J'/J$.
The first two parameters determine the orbital state, and the last one 
tunes the spin interactions. The magnetic order depends on all these
parameters as the superexchange due to $e_g$ electrons is 
{\it intrinsically frustrated\/} and contains both AF and FM terms
\cite{Fei99}, which become active for particular pairs of occupied 
orbitals on a given bond $\langle ij\rangle$, see Eq. (\ref{HJ}). 

Apart from the unphysical regime of negative $J'<0$ which enforces the
FM order independently of the orbital state, the AO order is necessary 
to stabilize it for $J'\sim 0$ and $|E_z|<0.5t$. This agrees with the 
Goodenough-Kanamori rules \cite{Goode}, and with the interplay between 
the magnetic and $e_g$ orbital correlations studied before in the 1D 
model \cite{Dag04}. 
The FM phase competes with the $G$-AF order stable for sufficiently 
large core spin superexchange $J'$ [Fig. \ref{fig:x=0}(a)]. However, the 
transition line depends also on the two remaining parameters: $E_z$ and 
$\kappa$. The dependence on $E_z$ is stronger --- for $E_z<-0.5t$ the AF 
order is stable already at $J'=0$, while for $E_z>0.5t$ the monolayer 
would be FM, if the AF superexchange $J'$ between core spins would not 
help to stabilize the AF order. This behavior is a manifestation of the 
large difference of 9:1 between the AF $e_g$ superexchange interaction 
\cite{Ole00}, depending on whether $e_g$ electrons are in $|x\rangle$ 
or in $|z\rangle$ orbitals, respectively. 
Altogether, the transition line between the FM and $G$-AF phase in only 
weakly dependent on $E_z$ in the regime of large $|E_z|>t$; once all 
electrons are found in either $|x\rangle$ or $|z\rangle$ ferro orbital
(FO) phase, further increase of $|E_z|$ does not change the state. 
Remarkably, the range of $C$-AF phase is very narrow and vanishes for 
$E_z>0.3t$, indicating that large overlap between
$|x\rangle$ orbitals plays a role in stabilizing this type of order,
which could also be a finite size effect. 

At the expected value of core spin superexchange $J'=0.025t$ in 
LaSrMnO$_4$, the $G$-AF state is found independently of $E_z$
[Fig. \ref{fig:x=0}(b)]. However, the experimental data suggest that
$|z\rangle$ orbitals are occupied in LaSrMnO$_4$ \cite{Sen05}, and we  
find that this situation is well described by $E_z\sim t$. When  
temperature increases to $\beta t=100$, the orbital state almost does 
not change and indeed the MC data for the density $n_z$ follow the line 
obtained at $T=0$ for $G$-AF phase, see Fig. \ref{fig:x=0}(b). Further 
increase of temperature promotes excitations to $|x\rangle$ orbitals
\cite{Sen05}, and this experimental finding is also reproduced by the 
present model \cite{pss06}. It is interesting to remark that the $G$-AF 
order is stabilized in this parameter 
regime by core spin AF superexchange $J'$, while the $e_g$ part of the
superexchange is FM, as the excitations from occupied 
$|z\rangle$ to unoccupied $|x\rangle$ orbitals dominate, with the 
hopping element larger by $\sqrt{3}$ than the one between two 
$|z\rangle$ orbitals, which would give instead the AF coupling. 

\begin{figure}[t!]
\includegraphics[width=6.5cm]{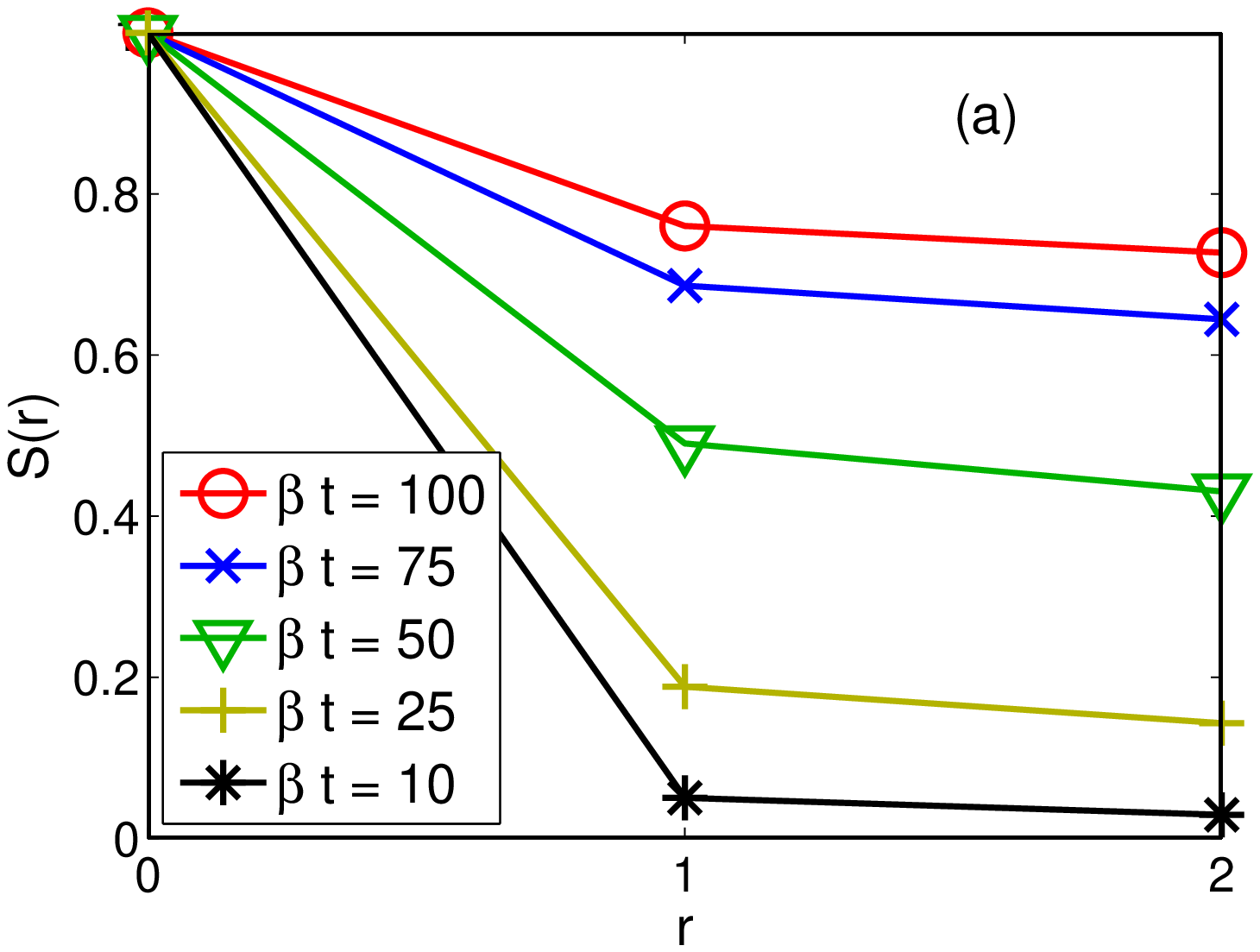}
\hskip .5cm
\includegraphics[width=6.6cm]{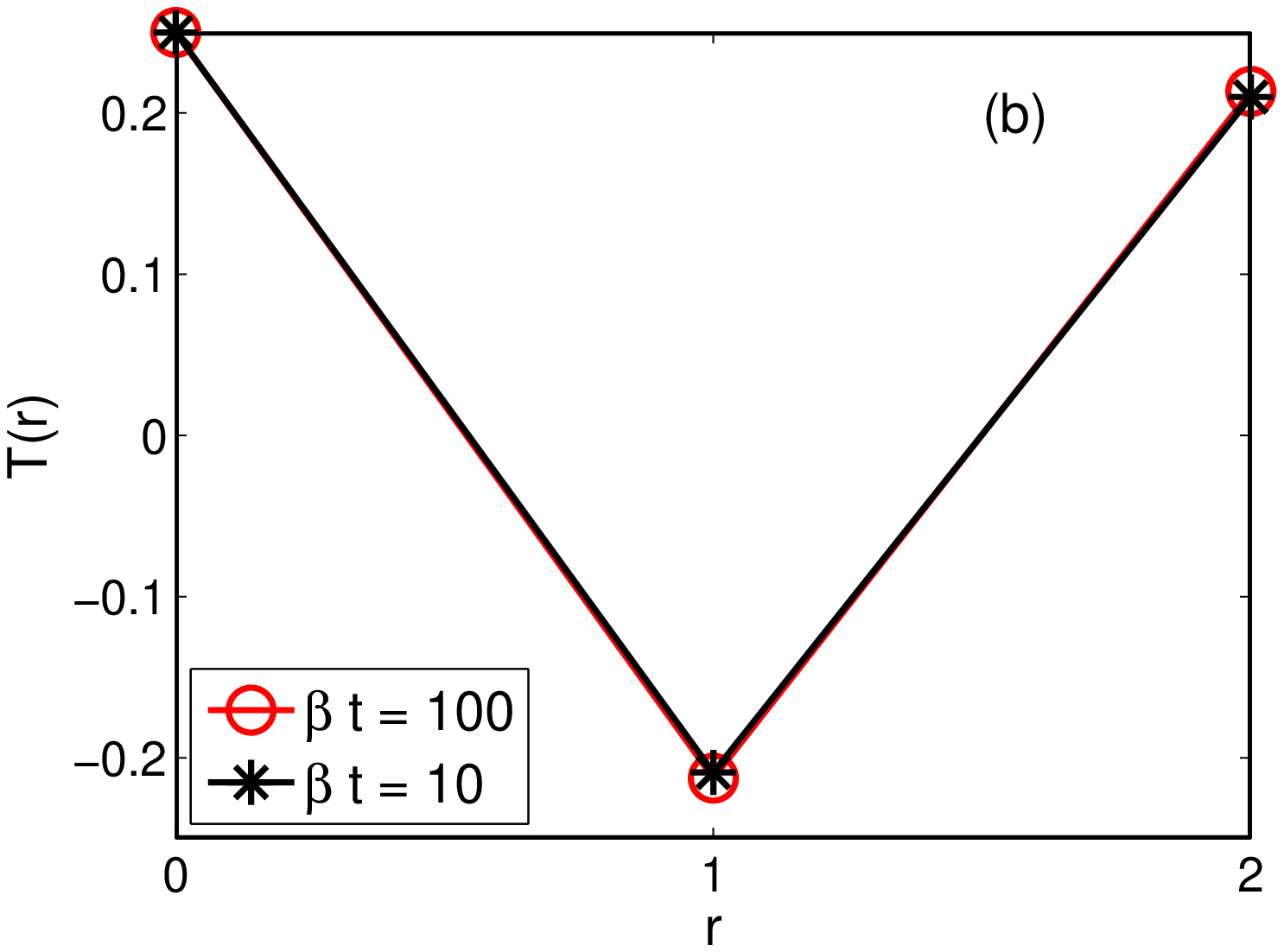}
\caption{(Color online)
Short range correlations for the undoped monolayer manganite ($x=0$) 
as obtained in the FM state with $\sqrt{8}\times\sqrt{8}$ cluster for 
increasing temperature:
(a) spin,
(b) orbital for $\theta=\frac{\pi}{2}$.
Parameters: $J=0.125t$, $\kappa=0.2t$, and $J'=0$.
}
\label{fig:sstt}
\end{figure}

At $J'=0$ the FM state is stable for $E_z>0$, but particularly when the 
electron densities in both $e_g$ orbitals are similar for $E_z\sim 0$, 
and the OO may form with occupied orbitals
\begin{equation}
\label{ao}
|\theta\rangle_{i\in A}=
 \cos\big(\textstyle{\frac{1}{2}}\theta\big)|z\rangle_i
+\sin\big(\textstyle{\frac{1}{2}}\theta\big)|x\rangle_i,  \hskip .7cm  
|\theta\rangle_{j\in B}=
 \cos\big(\textstyle{\frac{1}{2}}\theta\big)|z\rangle_j
-\sin\big(\textstyle{\frac{1}{2}}\theta\big)|x\rangle_j,
\end{equation}
on two sublattices $A$ and $B$. 
In this case the FM state is robust and rather large $J'\simeq 0.025t$ 
is necessary to switch the magnetic correlations to the $G$-AF state 
[Fig. \ref{fig:x=0}(a)]. This is also shown by the MC data for the 
electron density in $|z\rangle$ orbitals $n_z$, which remains 
practically unchanged from the ground state value still at temperature 
$\beta t =100$ [Fig. \ref{fig:x=0}(b)]. The density $n_z$ increases 
slower with $E_z$ than in the AF state, as the AO state (\ref{ao}) can 
form only when the electron density in $|z\rangle$ orbitals is similar
to that in $|x\rangle$ orbitals. Note also that a rapid drop of $n_z$
with decreasing $E_z$ near $E_z=-0.5t$ in the MC data for $J'=0$ 
accompanies the phase transition to the $G$-AF phase, see phase 
boundary in Fig. \ref{fig:x=0}(a).
 
We emphasize that, similar to LaMnO$_3$ \cite{Fei99}, the AO state with 
FM spin order is stabilized not only by the orbital superexchange terms 
in Eq. (\ref{HJ}), but also by the orbital JT interactions (\ref{HJT}). 
This results in the separation of the energy scales at finite  
$\kappa=0.2t$ shown in Fig. \ref{fig:sstt} --- the magnetic order is 
lost first in the range of temperature $\beta t\sim 50$, while the AO 
order is more robust and starts to weaken only above $\beta t\sim 10$.
While at $\kappa=0$ the orbital correlations decrease somewhat faster 
with increasing temperature \cite{pss06}, they are still quite 
pronounced when the magnetic order is lost. This shows that FM 
correlations would be typically lost at lower temperature $T\sim 150$ K
than the orbital correlations, to a large extent due to conflicting 
trends in magnetic interactions which compete with each other. 

\begin{figure}[t!]
\includegraphics[width=6.7cm]{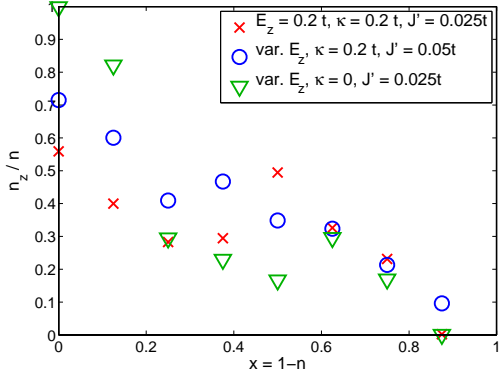}
\caption{(Color online)
Percentage of electrons found in $|z\rangle$ orbitals $n_z/n$ as 
a function of doping $x$ obtained by MC on a $\sqrt{8}\times\sqrt{8}$ 
cluster for three parameter sets: 
($i$) $E_z=0.2t$, $\kappa=0.2t$, $J'= 0.025t$, 
($ii$) variable $E_z=\frac12(1-x)$, $\kappa=0.2 t$, $J'=0.05t$, and 
($iii$) variable $E_z=(\frac12-x)$, $\kappa=0$, $J'=0.025t$  
(the last set of data has also been shown in Ref. \cite{Dag06}).
Parameters:  $J=0.125t$, $V=t$.
}
\label{fig:nz}
\end{figure}

In order to understand the evolution of magnetic interactions with 
increasing hole doping, it is instructive to consider first the orbital 
occupation. The geometry of the monolayer compounds suggests positive
crystal field splitting $E_z>0$ of $e_g$ orbitals, so the $|z\rangle$ 
orbitals which stick out of the $ab$ plane are favored. We have found 
that for reasonable positive values of $E_z$ the majority of electrons 
is still in $|z\rangle$ orbitals at low doping, see Fig. \ref{fig:nz}. 
However, for $x>0.2$ both $e_g$ orbitals are either nearly equally 
populated, or the $|x\rangle$ orbitals dominate, i.e., $n_x>n_z$. 
This behavior confirms the trend found before in the 2D model at orbital 
degeneracy \cite{Mac99}. Here we show that the FO order 
with $|x\rangle$ orbitals occupied cannot form as long as crystal field 
splitting is not lower than $E_z\sim 0$. This helps to suppress the 
kinetic energy in $ab$ plane, so the double exchange interaction is 
weak at finite doping. Actually, for this reason the FM state cannot 
form in the range of intermediate doping $0.2<x<0.4$ when $E_z>0$ and 
the core spin superexchange is in the expected range, $J'\sim 0.05t$. 
Altogether, we argue that the orbital liquid state which forms only in 
cubic geometry \cite{Fei05} plays an essential role in stabilizing the 
FM state. 

We believe that the crystal field splitting gradually decreases with 
increasing $e_g$ hole density, as the distortions of MnO$_6$ octahedra 
are then expected to decrease and could vanish in the limit of $x=1$. 
Qualitatively this situation is modeled by variable $E_z=\frac12(1-x)t$ 
in Fig. \ref{fig:nz}. While about 70\% of electrons fill then 
$|z\rangle$ orbitals at $x=0$, the filling of these orbitals at $x=0.5$ 
is still about 35\% (for $E_z=\frac14 t$). In contrast, when $E_z=0$
(last set of data in Fig. \ref{fig:nz}), the filling of $|z\rangle$ 
orbitals drops below 20\%, being too low to form the observed OO state
\cite{Wil05}. Below we analyze
the influence of the electron density distribution and the microscopic 
parameters on the stability of the CE-AF phase at $x=0.5$. 

\begin{figure}[t!]
\includegraphics[width=6.6cm]{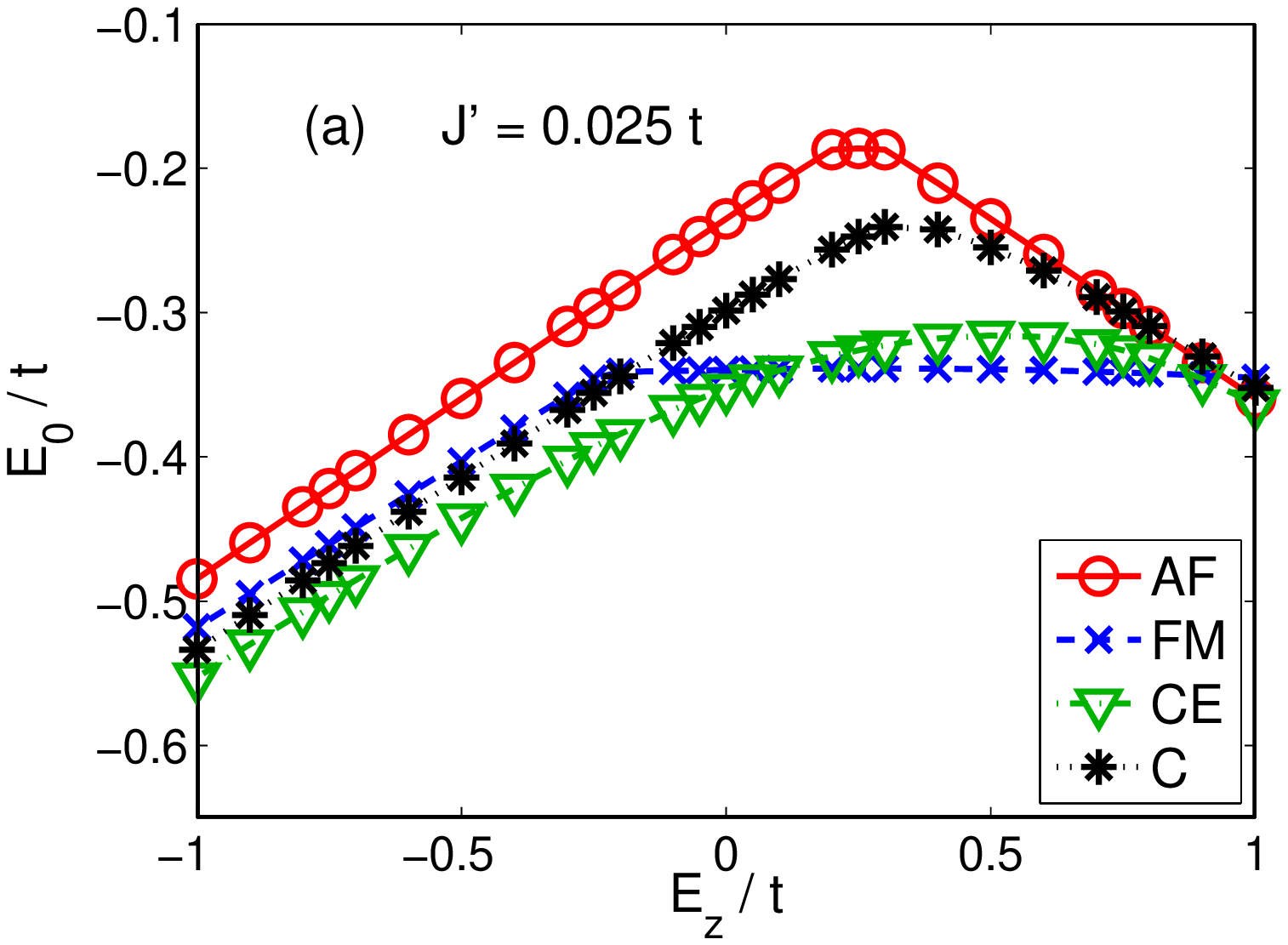}
\hskip .5cm
\includegraphics[width=6.6cm]{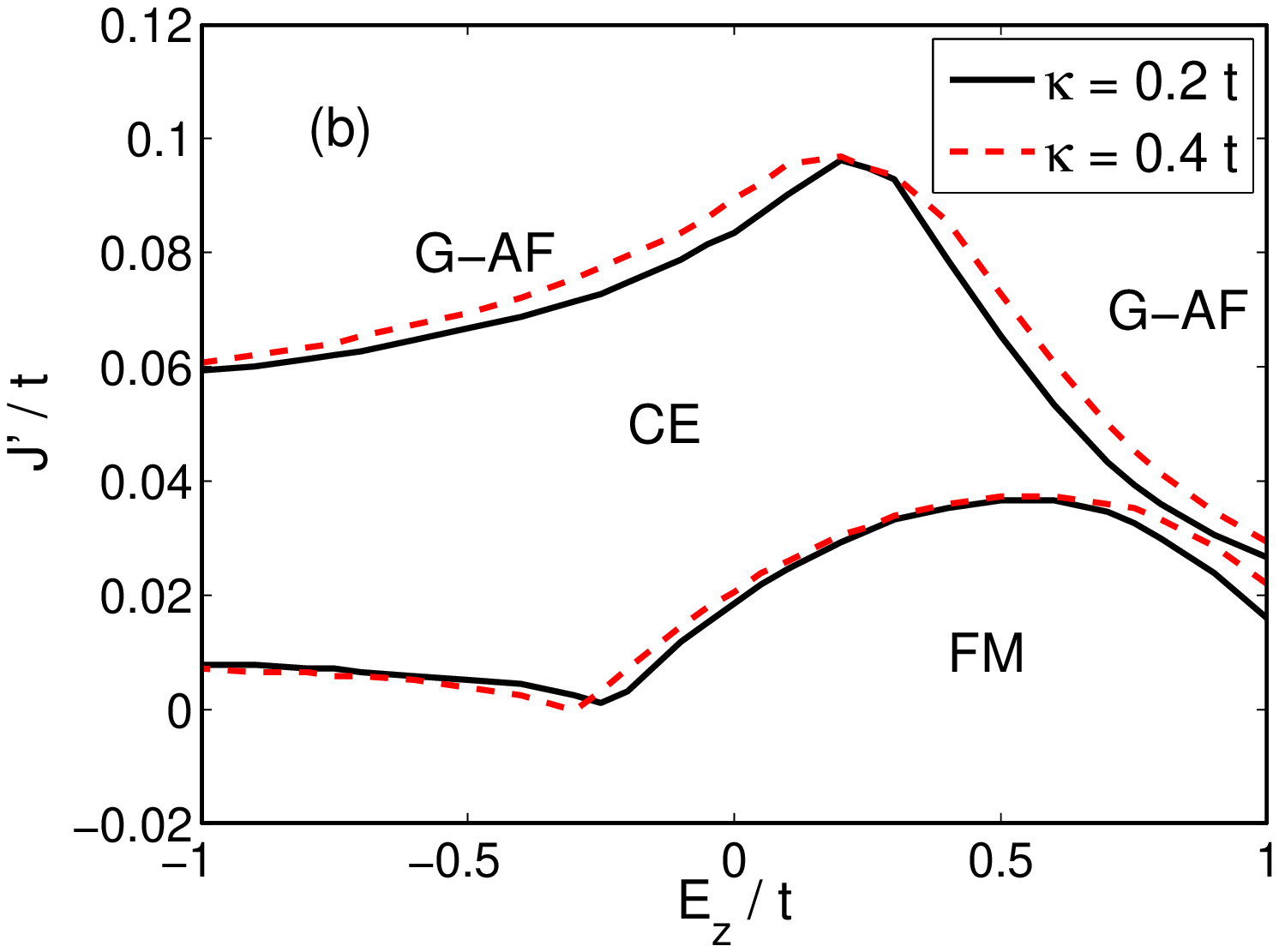}
\caption{(Color online)
Evolution of the magnetic ground state for $x=0.5$ doping as obtained 
with an $\sqrt{8}\times\sqrt{8}$ cluster for increasing crystal field 
splitting $E_z$ of $e_g$ orbitals: 
(a) energies for various possible magnetic phases ($G$-AF, FM, CE-AF, 
    $C$-AF) for $\kappa=0.2t$ and $J'=0.025t$;
(b) phase diagram in $(E_z,J')$ plane for $\kappa=0.2t$ (solid lines)
    and $\kappa=0.4t$ (dashed lines).   
Parameters: $J=0.125t$, $V=t$. 
}
\label{fig:phdez}
\end{figure}

The CE phase was observed in a monolayer La$_{0.5}$Sr$_{1.5}$MnO$_4$ 
compound \cite{Lar05}, corresponding to half doping ($x=0.5$). This 
type of order is quite exotic and there were several attempts in the 
literature to identify the leading parameters which stabilize this 
complex type of coexisting charge, orbital, and magnetic order. We
have performed extensive ground state ($T=0$) calculations using 
$\sqrt{8}\times\sqrt{8}$ clusters with periodic boundary conditions,
which are large enough to capture the essential interactions deciding
about the nature of the ground state \cite{Bal04}, as we also concluded 
recently \cite{Dag06}. 

\begin{figure}[b!]
\includegraphics[width=6.7cm]{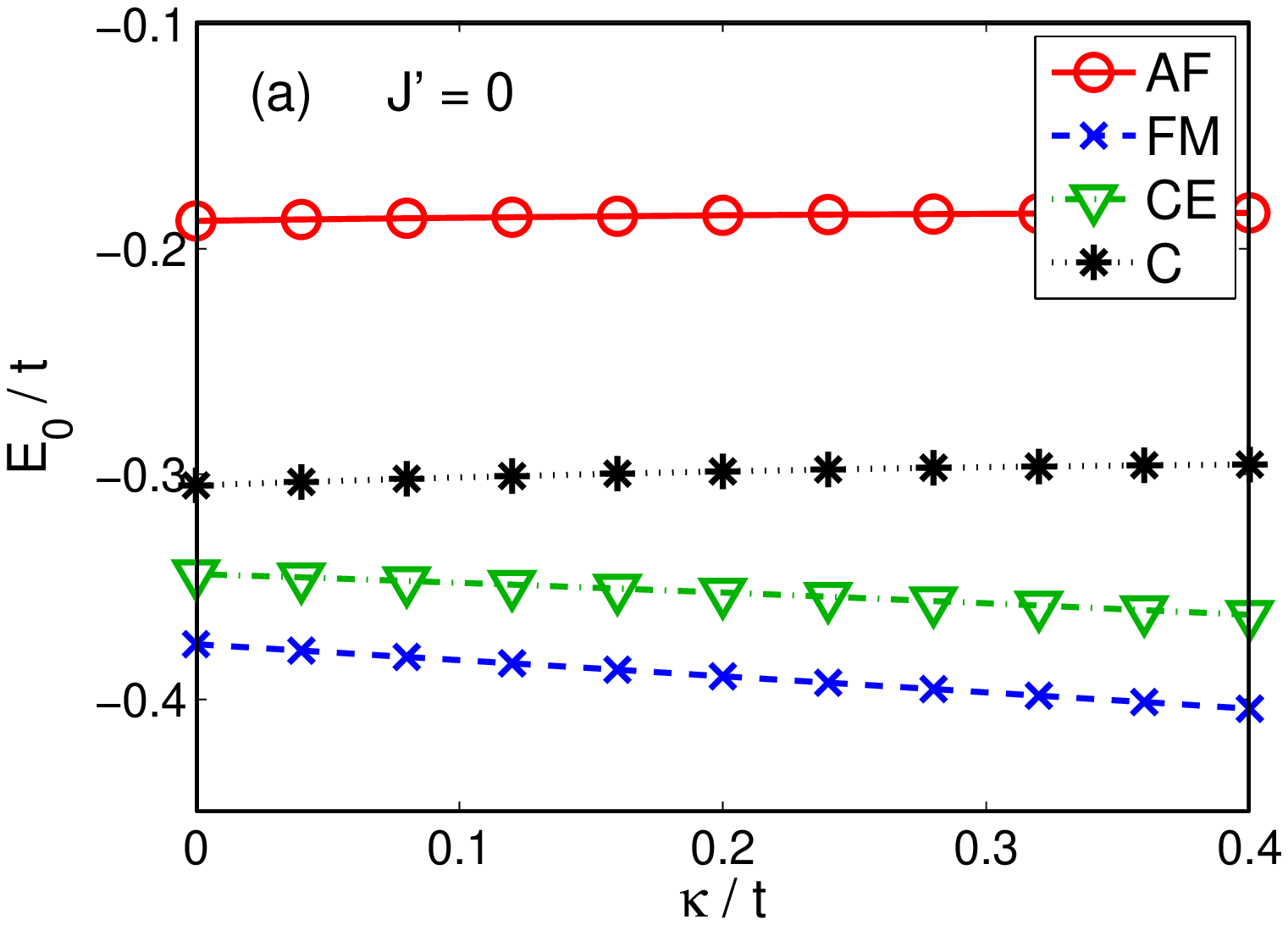}
\hskip .5cm
\includegraphics[width=6.6cm]{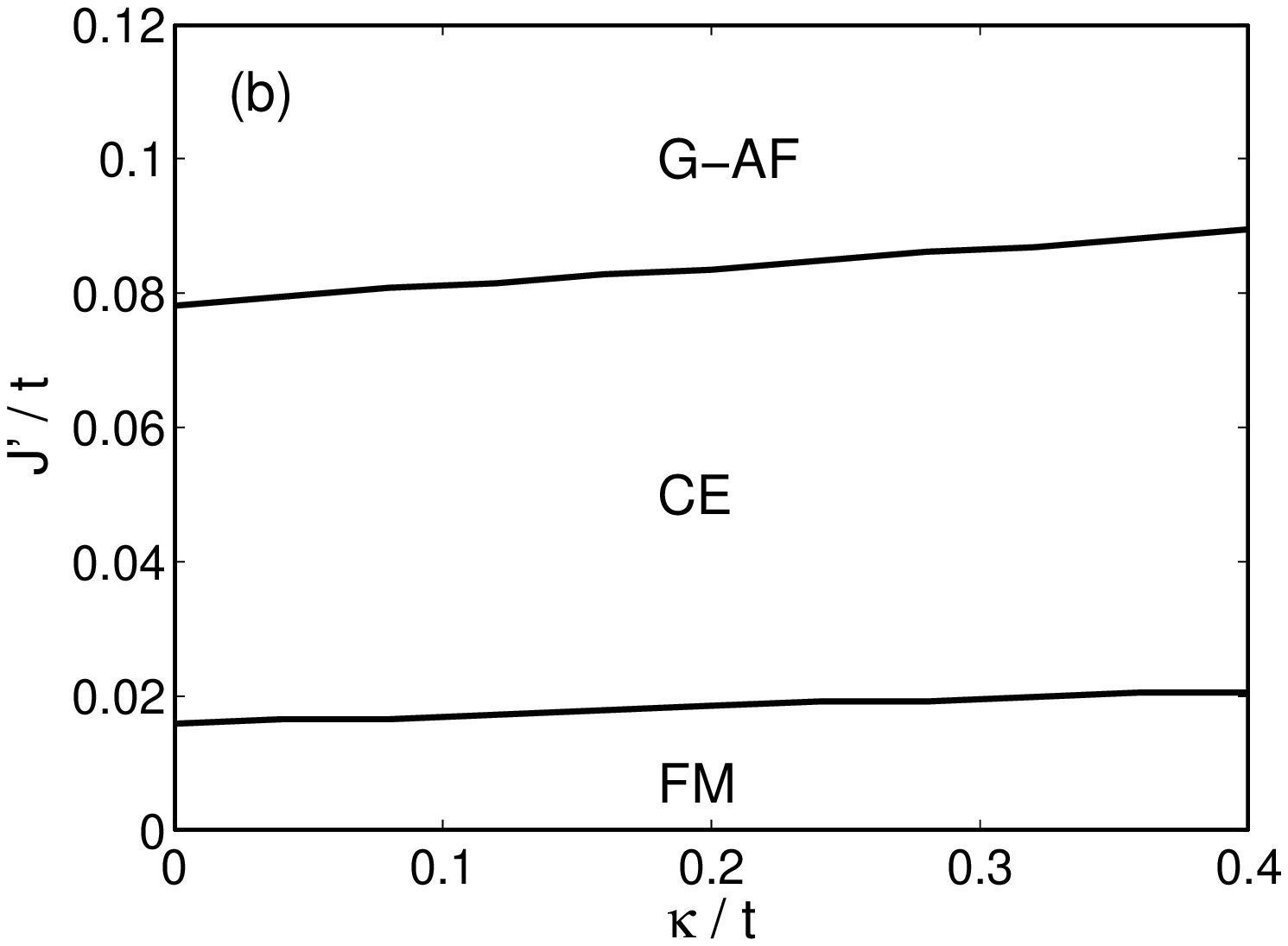}
\caption{(Color online)
Magnetic phases for increasing JT interaction $\kappa$ as obtained with 
an $\sqrt{8}\times\sqrt{8}$ cluster for $x=0.5$: 
(a) ground state energies of various phases ($G$-AF, FM, CE-AF, $C$-AF) 
    for $J'=0$;
(b) phase diagram in $(\kappa,J')$ plane.
Parameters: $J=0.125t$, $E_z=0$, and $V=t$. 
}
\label{fig:phdkappa}
\end{figure}

In the absence of JT interactions (at $\kappa=0$), the CE phase arises 
as a compromise between competing AF and FM interactions, roughly 
speaking in a range of $0<J'<0.1t$ and $V<1.5t$ \cite{Dag06}.  
In contrast to the undoped case, the magnetic interactions at half 
doping are in general anisotropic and the orbital state can be then 
tuned to support FM spin order along a certain direction by double 
exchange mechanism, while the superexchange is predominantly AF in 
the other direction. In this way the $C$-AF (studied before in the 
ladder geometry \cite{Neu06}) and CE-AF phase may arise and strongly 
compete with each other due to rather similar kinetic and interaction 
energies \cite{Dag06}, as long as $\kappa=0$. The situation changes 
however when the JT interactions are considered, as finite 
$\kappa=0.2t$ used for the data of Fig. \ref{fig:phdez}(a) favors also 
alternation of two quasi-planar orbitals,  
\begin{equation}
\label{ceoo}
|+\rangle_{i\in A}=
 \textstyle{\frac{1}{\sqrt{2}}}\big(|z\rangle_i+|x\rangle_i\big)
\hskip .7cm  
|-\rangle_{j\in B}=
 \textstyle{\frac{1}{\sqrt{2}}}\big(|z\rangle_i-|x\rangle_i\big),
\end{equation}
on the charge majority sites in the regime of $0<E_z<0.5t$, where 
electron densities in $|x\rangle$ and $|z\rangle$ orbital are similar, 
rather than directional $3x^2-r^2/3y^2-r^2$ orbitals. For these 
parameters the $C$-AF phase is destabilized. Thus, for a moderate 
value of core spin superexchange $J'=0.025t$, the CE phase competes only
with the FM phase, and the latter one takes over for $E_z>0.15t$, i.e., 
when the population of both $e_g$ orbitals is almost equal and the 
double exchange energy dominates. This is also seen in the phase diagram 
of Fig. \ref{fig:phdez}(b) --- increasing $E_z$ pushes the CE phase to 
higher values of $J'$ in a range of $-0.25t<E_z<0.5t$. Also the 
transition between the CE and $G$-AF occurs at somewhat higher values of 
$J'$ when $E_z$ increases in a range of $-t<E_z<0.25t$. 
Quite remarkably, when $E_z$ increases beyond these limits, an opposite 
trend is found, as then only $|z\rangle$ orbitals are occupied and the 
OO which accompanies the CE phase (\ref{ceoo}) cannot develop. 
Therefore, the CE phase collapses, and may exist only in a very narrow 
window of $J'$ values in between the FM and $G$-AF phase.   

\begin{figure}[t!]
\includegraphics[width=6.7cm]{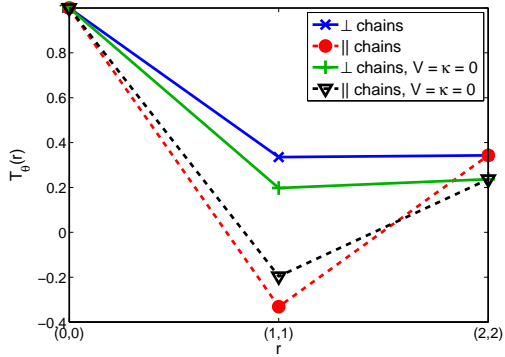}
\caption{(Color online)
Orbital correlations along the diagonal $(1,1)$ and $(1,\bar{1})$ 
directions in the CE phase ($x=0.5$) for $J=0.125t$, and for: 
$V=t$, $\kappa=0.2t$ (solid lines), and 
$V=\kappa=0$ (dashed lines).
}
\label{fig:ttx=1/2}
\end{figure}

Surprisingly, the dependence of all magnetic energies on the value of 
the JT interaction $\kappa$ is quite weak, see Fig. \ref{fig:phdkappa}.
This interaction stabilizes the OO both in the FM and in CE phase, so 
the energies of these phases decrease with increasing $\kappa$  
[Fig. \ref{fig:phdkappa}(a)]. However, the actual boarder lines between
these two phases and between the CE and $G$-AF phase, respectively, are
determined by the value of $J'$ [Fig. \ref{fig:phdkappa}(b)], and depend 
also on the electron density distribution over $|x\rangle$ and 
$|z\rangle$ orbitals, as discussed above. 

\begin{figure}[b!]
\includegraphics[width=6.1cm]{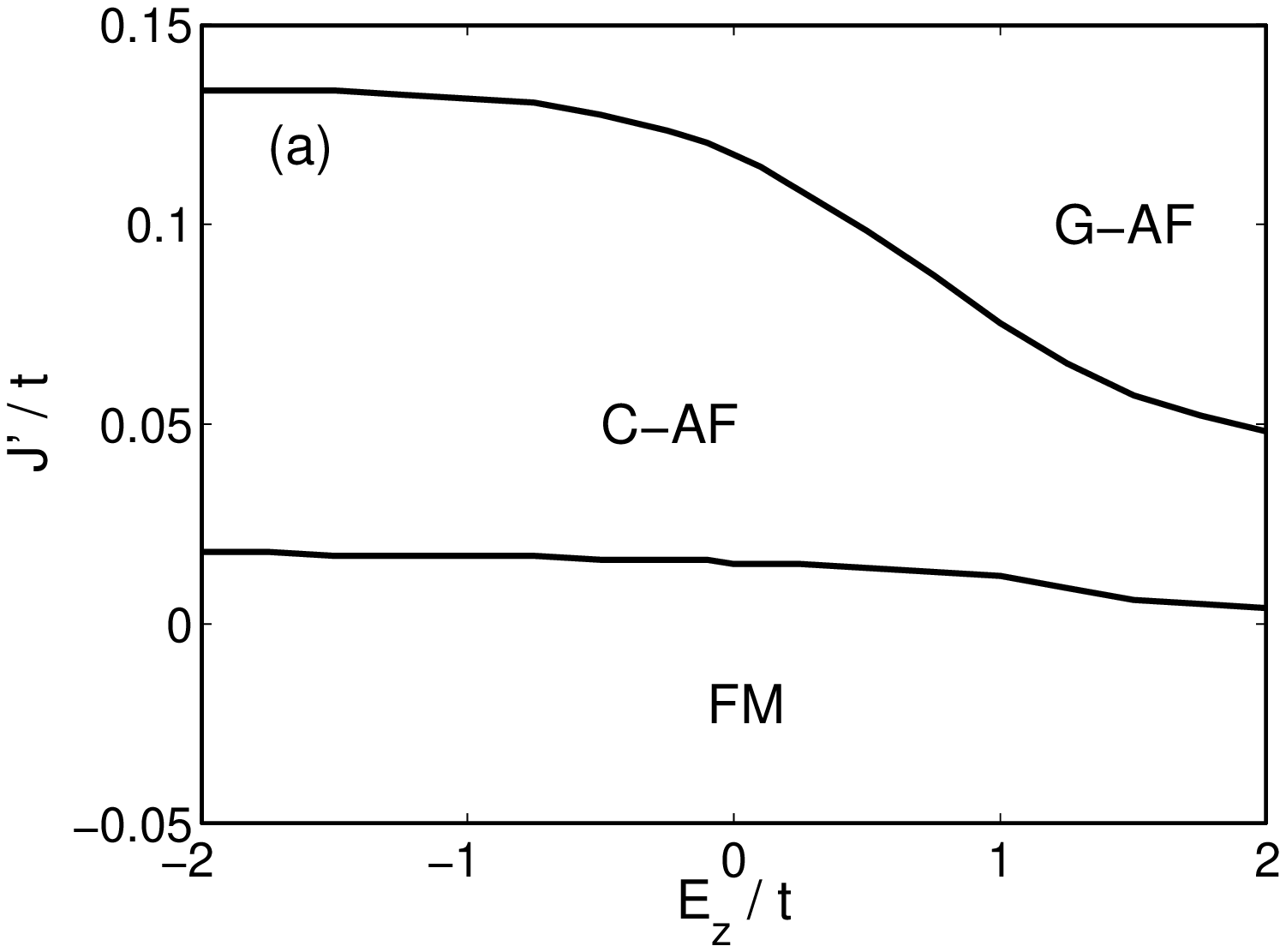}
\hskip .7cm
\includegraphics[width=7cm]{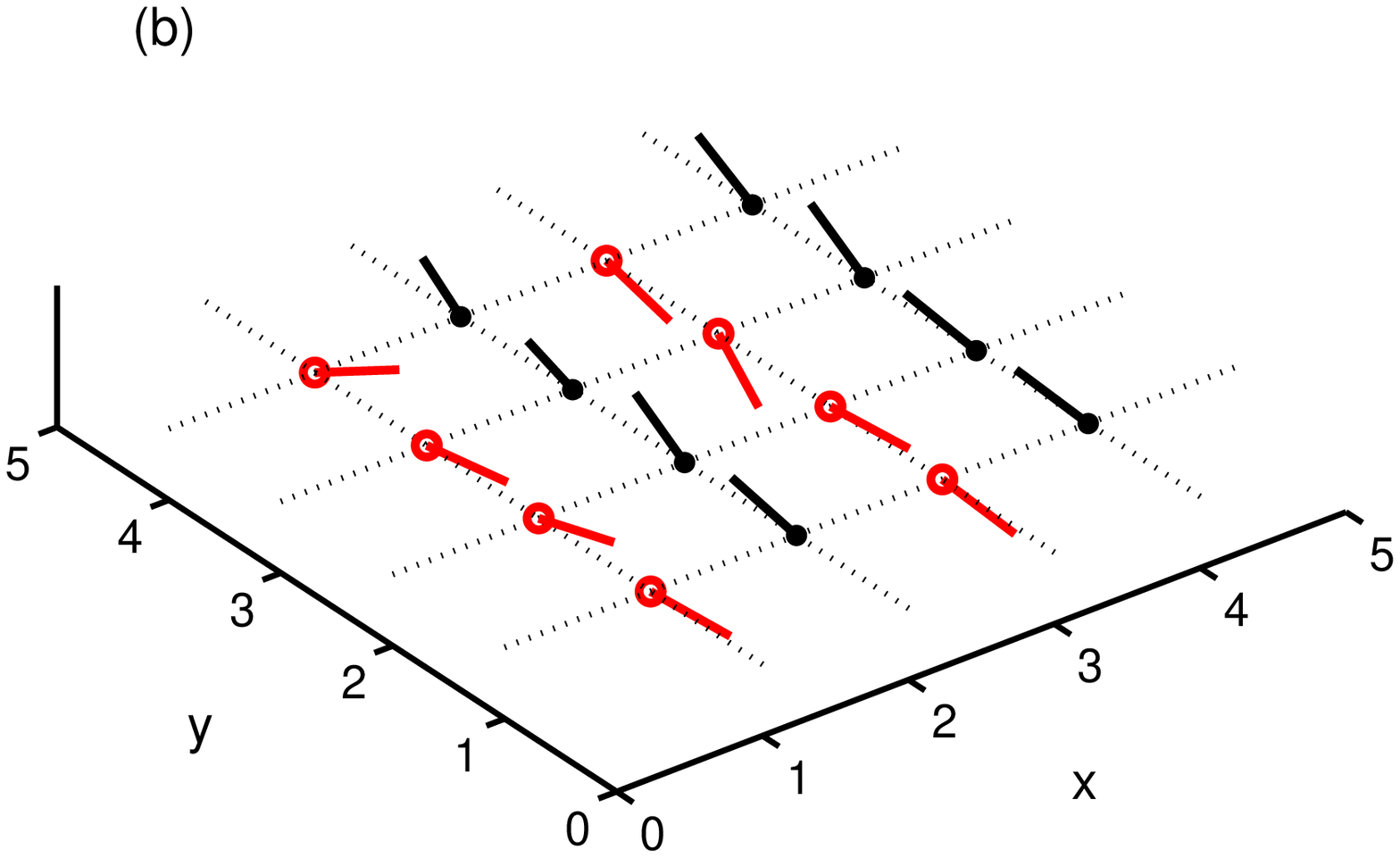}
\caption{(Color online)
Competition between AF and FM spin interaction at $x=0.75$ doping,
as obtained with a $4\times 4$ cluster:
(a) phase diagram in $(E_z,J')$ plane with FM, $C$-AF and $G$-AF phases;
(b) MC snapshot of the $C$-AF phase for $E_z= t$, $J'=0.05t$, 
    $\beta t=100$.
Other parameters: $J=0.125t$, $\kappa=0.2t$, and $V=t$, .
}
\label{fig:x=3/4}
\end{figure}

We also note that a finite value of $V\sim t$ is important and helps to 
stabilize the checkerboard charge order coexisting with the OO in the 
CE phase. As expected in this state, the $|+\rangle$ and $|-\rangle$ 
occupied orbitals (\ref{ceoo}) alternate on the sites with increased 
electron density when $V=t$ and $\kappa=0.2t$ (Fig. \ref{fig:ttx=1/2}). 
This OO is well developed, but still rather far from the idealized one, 
as $e_g$ electrons are partly delocalized. We note that this partial 
delocalization is crucial for the double exchange which favors the 
FM order along the zigzag chain. 

Although the competition between the AF superexchange and the FM double
exchange persists also at higher doping $x>0.5$, the checkerboard 
charge ordered state is destabilized, and the optimal conditions for 
the CE phase are not fulfilled. Instead, one finds a robust tendency 
towards the $C$-AF phase, as we demonstrate by the data obtained with 
a $4\times4$ cluster for doping $x=0.75$ in Fig. \ref{fig:x=3/4}(a).
As for $x=0.5$, the FM phase is stable at $J'=0$. In this case the 
density of $e_g$ electrons is so low that the AF superexchange 
between Mn$^{3+}$ and Mn$^{4+}$ ions is rather weak and the $C$-AF 
phase, observed experimentally in Nd$_{1-x}$Sr$_{1+x}$MnO$_4$ 
\cite{Kim02}, is obtained only when $J'>0.02t$, i.e., in the expected 
range obtained by analyzing the $t_{2g}$ superexchange. The value of 
critical $J'$ 
decreases somewhat with increasing $E_z$, as the double exchange 
weakens when $|z\rangle$ orbitals are occupied for positive values 
of $E_z$. This effect is quite strong for the second phase boundary, 
and one finds that the $C$-AF phase survives in a considerably broader
range of $J'$ when $E_z<0$. Actually, the critical value of $J'$ 
exchange changes roughly by a factor of three between $E_z=-2t$ and 
$E_z=+2t$, which originates from the strength of the double exchange
proportional to the respective hopping elements between two $|x\rangle$
or two $|z\rangle$ orbitals, respectively.

\section{Discussion and summary}

The present study clarifies that orbital degrees of freedom are of
crucial importance for a complete understanding of magnetic correlations 
in monolayer manganites. We treated a realistic orbital $t$--$J$ model 
including Coulomb and JT interactions and investigated charge, 
spin and orbital intersite correlations in monolayer manganites. 
The obtained results revealed a close relationship between orbital and 
magnetic order which follows the classical Goodenough-Kanamori rules 
at $x=0$ \cite{Goode}. The magnetic phases found in different doping 
regimes, where double exchange also contributes, are in accordance with
experiments over the entire doping range $0\leq x<1$.

For the undoped monolayer LaSrMnO$_4$ manganite the model predicts 
$G$-AF order, when the crystal field splitting of $e_g$ orbitals 
$E_z\sim t$ and the core spin superexchange $J'\sim 0.03t$ are in the
expected range. Due to large crystal field splitting the role played 
here by the intersite orbital interactions $\propto\kappa$ is in this 
case negligible. We emphasize that the correct treatment of electron 
correlation effects within the orbital $t$--$J$ model \cite{vdB00} is 
crucial for the qualitatively correct description of the magnetic state 
in LaSrMnO$_4$, and quite different results are obtained when strong 
on-site Coulomb repulsion $U$ is neglected \cite{Hot03}.

Another success of the model is that it predicts the CE phase at half
doping with physically realistic parameters for layered manganites,
i.e., for {\it small\/} $t_{2g}$ superexchange $J'\sim 0.03t$, as 
deduced \cite{Ole05} from the analysis of exchange constants in 
LaMnO$_3$, rather than for unrealistically large $J'>0.2t$ 
\cite{Efr04}. The orbital interactions promoted by the JT effect play 
here a very important role in stabilizing this phase, as they support 
precisely the type of the AO order on the checkerboard lattice which 
promotes FM interactions along the zigzag chains by double exchange 
mechanism. The occupied orbitals are closer to the planar 
$z^2-x^2/y^2-z^2$ orbitals \cite{Wil05}, rather than to the directional 
$3x^2-r^2/3y^2-r^2$ orbitals, which could also be explained by a 
crystal field $E_z>0$ \cite{Dag06}, as is frequently believed. Also at 
doping $x\simeq 0.75$, the present orbital $t$--$J$ model predicts the 
$C$-AF phase in agreement with experiment \cite{Kim02}.

Summarizing, the present study shows that the magnetic phase diagrams
of monolayer manganites demonstrate a competition between different 
types of order which follows from coexisting FM and AF terms in the 
spin superexchange, and the FM double exchange at finite hole doping. 
At the magnetic transitions the orbital order changes simultaneously,
as it supports particular types of magnetic order, in agreement with 
the Goodenough-Kanamori rules. We find it quite remarkable that yet
another application of the ideas developed about thirty years ago 
along the derivation of the celebrated spin $t$--$J$ model \cite{Cha77}
turned out to be so successful in monolayer manganites, and reproduced 
several generic features observed experimentally in this class of 
strongly correlated electron systems.

\acknowledgments
We thank Peter Horsch for insightful discussions.
A.~M.~Ole\'s acknowledges support by the Polish Ministry
of Science and Education under Project No. N202 068 32.

\end{document}